\documentclass{ws-procs9x6}
\begin{document}
\title{Color transparency: 33 years and still running. }
\author{ M.~STRIKMAN} 
\address{ Dept.\ of Physics, Pennsylvania State University, 
University Park, PA 16802, USA}
\begin{abstract}I review history of the color transparency (CT)  which started with discovery of the $J/\psi$ meson,  discovery  of high energy CT phenomena and the recent progress  in the investigations of CT at intermediate energies.
\\
\end{abstract}
%
%
\vspace{1cm} 

\section{Historical introduction}
One of the distinctive properties of QCD is the suppression of the interaction of small size color singlet wave packets with hadrons. It plays a key role in ensuring approximate Bjorken scaling in deep inelastic scattering, in proving QCD factorization theorems for high energy hard exclusive processes, etc. It leads to a number of color transparency (CT) phenomena in the hard coherent / quasielastic interactions with nuclei  at high energies. Also, the CT phenomenon allows to probe minimal small size components in the hadrons. In addition,  at intermediate energies CT phenomena provide   unique probes of the space time evolution of wave packets which is relevant for interpretation of the  RHIC  heavy ion collision  data.

For me the story of CT goes back to the discovery of $J/\psi$. It was impossible to explain within the 
concepts of the pre-QCD theory of strong interactions why the decay width of $J/\psi$ is so small, and (this was learned soon after its discovery) why the photoproduction cross section is so small. These issues were subject of numerous discussions between Leonya Frankfurt and Volodya Gribov during the winter of 74-75 with VG trying to reconcile $J/\psi$ properties with the soft Pomeron logic and  LF arguing that for a system consisting of heavy quarks the radius should be significantly smaller than one given by radius of pion emission (this was in contrast to the widely accepted  idea at that time    due to Fermi, that   the radius of a hadron is determined by the pion cloud and therefore should be approximately universal). More generally LF argued that all matrix elements involving heavy quarks should  be suppressed, leading to a strong reduction of the cross section of $J/\psi$ - nucleon interaction ($\propto 1/M_{J/\psi}^2$) and "an unusual conclusion that nucleon becomes transparent to the hadrons built of heavy quarks" 
\cite{Frankfurt:1975sy}. This was a clear break with the strong interaction picture with one soft scale which was discussed before $J/\psi$.

A perturbative model for the  interaction of hadrons via two gluon exchange was applied to $J/\psi -N $ interaction by  Gunion and Soper
\cite{Gunion:1976iy} who demonstrated that within the model the smallness of the $J/\psi$-nucleon interaction is related to the spacial 
small size of $J/\psi$.  Arguments that the suppression should be  present also  in the nonperturbative domain were given in \cite{Frankfurt:1985cv} where it was argued also that small $J/\psi (\psi^{\prime})$ nucleon cross section extracted 
from the photoproduction data
using the vector dominance model  underestimates the genuine $J/\psi - N$ and especially $\psi^{\prime} - N$ cross section by a large factor.

An independent development was the discussion of the hard exclusive processes like nucleon form factor, large angle hadron-hadron scattering in the  large $Q^2$ limit.  
A debate was going on whether the minimal Fock space 
components highly localized in space give the dominant contribution in the kinematic range studied experimentally, or the process is dominated by the end point contributions corresponding to quark - gluon configurations of average size. For a recent review see \, \cite{Radyushkin:2004sr} .

A. Mueller has suggested to use exclusive processes off nuclei, namely large angle reaction $pA\to pp (A-1)$ in order to discriminate between the two mechanisms \cite{Mueller82}, while S.Brodsky\cite{Brodsky82}
  made a prediction that the cross of the process  $\pi A\to \pi p (A-1)$ should be proportional to the number of protons in the target.
It is feasible to study these processes  as well as   quasielastic electron - nucleus scattering only in the kinematics where at least one  hadron in the final state has relatively small momentum leading to a need to take into account space time evolution of the quark-gluon wave packets involved in the collision
 which greatly reduces the CT effect \,  \cite{FLFS88} .
 
 This called for finding
  high energy processes which are dominated by the interaction of hadrons in small size configurations which could be legitimately calculated in pQCD and which are not affected
by the space-time evolution of small wave packets. A key observation was   that,  due to the  possibility of treating configurations as frozen during the collision process one can introduce a notion of the cross section of scattering of a small dipole configuration (say $q\bar q$) of transverse size $d$ on the nucleon \, \cite{Frankfurt:1993it, Blaettel:1993rd} which in the leading log approximation is given by \, \cite{Frankfurt:2000jm}
\begin{equation}
\sigma(d,x)= {\pi^2\over 3} \alpha_s(Q^2_{eff}) d^2\left[x_NG_N(x,Q^2_{eff}) +2/3 x_NS_N(x_N, Q^2_{eff})\right],
\label{pdip}
\end{equation}
where $Q^2_{eff} = \lambda/d^2, \lambda= 4 \div 10$, and  $S$ is the  sea quark distribution  for quarks making up the dipole.  Here, in difference from the original estimate, we include also the contribution of quark exchanges which is important for the interactions  at intermediate energies. Note that Eq.(\ref{pdip}) predicts a rapid  increase of the dipole -hadron cross section with increase of energy which is qualitatively  different from the expectation of the two gluon exchange model \, \cite{Gunion:1976iy} where cross section does not depend on energy.

First, we will consider the case a more simple case of high energy CT where only two conditions are required: dominance of small size configurations and smallness of $q\bar q - N $ interaction. Next we will consider a 
more complicated case  of  CT in the     intermediate energy processes where it is masked to large extent by the expansion effects. For a recent extensive review of the CT phenomena see \cite{Miller:2007zzd}.

\section{Discovery of high energy CT}
To observe CT in a high energy process one needs to find a trigger which selects  small size configurations in the projectile.
One idea is to select a special final state: diffraction of a pion into two high transverse momentum jets. 
Qualitatively one expects in this case $d\sim 1/p_t(jet)$. Another idea is to select a small initial state - diffraction of a longitudinally polarized virtual photon into a meson. It employs the decrease of the transverse separation between $q$ and $\bar q$ in the wave function of $\gamma^*_L, d\propto 1/Q$. The pQCD results for these processes where first derived in 
\, \cite{Frankfurt:1993it,Brodsky:1994kf} , with the proofs of the QCD factorization for these processes given for dijet production in \, \cite{Frankfurt:2000jm} and for meson production in \, \cite{Collins:1996fb} (where in addition to production of vector mesons a general case of meson production: $\gamma^*_L + N \to "meson \, system" + "baryon\, system" $ was considered).

\subsection{Pion dissociation into two jets}
The space time picture of the process is as follows - long before the target pion fluctuates into $q\bar q$ configuration with transverse separation $d$, which elastically scatters off the target with an amplitude which for $t=0$ is given by Eq.(\ref{pdip})  (up to 
small corrections due to different off shellness of $q\bar q$ pair in the initial and final states, followed by the transformation of the pair into two jets. A slightly simplified final answer is
\begin{equation}
A(\pi\,  N \to \, 2\, jets\, + \,N) (z,p_t,t=0) \propto \int d^2d \psi^{q\bar q}_{\pi}(d,z) \sigma_{q\bar q - N(A)}(d,s)e^{ip_td}, 
\label{dijet}
\end{equation}
where $z$ in the light-cone fraction of the pion momentum carried by a quark, $\psi^{q\bar q}_{\pi}(z,d)\propto z(1-z)_{d\to 0}$ is the  quark-antiquark Fock component of the meson light cone wave function. Presence of the plane wave factor in the final state leads to an expectation of an earlier onset  of  scaling than in the case of the vector meson production where vector meson wave function enters.

The FNAL experiment \cite{Aitala:2000hc} confirmed key CT predictions
of \, \cite{Frankfurt:1993it} : a) a strong increase of   the cross section of the $\pi +A \to "two \, jets" + A$ process with A(A=carbon, and platinum):  $\sigma \propto A^{1.61\pm 0.08}$ as compared to the prediction $\sigma \propto  A^{1.54}$
\footnote{In QCD a naive expectation of the  CT that the amplitude is proportional to A is modified \, \cite{Frankfurt:1993it,Brodsky:1994kf} due to the leading twist gluon shadowing which should be present at  sufficiently small x. This effect is not important for the x range of the experiment \cite{Aitala:2000hc} .}, b) the $z^2(1-z)^2$ dependence of the cross section on the fraction of energy $z$ carried by the jet, c) the 
$k_t$ dependence of the cross section. Note that the CT prediction for the A-dependence was a factor of seven different from the A-dependence  for the soft diffraction. 

In the recent update of the analysis Ashery reported \cite{Ashery:2005wa}  a fit to the $z$ distribution using Gegenbauer polynomials for different ranges of $p_t$. For $1.25\le p_t\le 1.5 \, $GeV/c higher order polynomials appear to be important. Since the CT is observed for this $p_t$ range  as well this indicates that squeezing occurs already  before the leading term $(1-z)z$ dominates. 

\subsection{Vector meson production at HERA}

Exclusive vector meson production
was extensively studied at HERA. 
The leading twist picture of the process\cite{Brodsky:1994kf}   is,  in a sense, a mirror image of  the dijet production  - virtual longitudinally polarized photon first transformed to a small transverse size pair which interacts elastically with a target and next transforms to a vector meson. Hence the process is described by the same equation (\ref{dijet})
as for pion case with a substitution of the plane wave $q\bar q$ wave function by the $q\bar q$ wave function of the longitudinally polarized virtual photon. 

The extensive studies of the vector meson production were performed at HERA.  Several of the theoretical predictions were confirmed including fast  x-dependence of the process at large $Q^2$, consistent with the x-dependence of $G_N^2(x,Q_{eff}^2)$, and convergence of the t-dependence to the universal one at large $Q^2_{eff}$ where it is given by the two gluon form factor.
At the same time the data  confirm a conclusion of the model studies \,  \cite{Frankfurt:1995jw} that in a wide range of virtualities one needs to take into account a higher twist effect of the finite transverse size of $\gamma_L$  to explain the absolute cross section and t-dependence of the data. The  leading twist dominance  for the absolute cross section for all mesons and for the t-dependence for light mesons requires very large $Q^2$ since only in this case one can neglect the transverse size of the $q\bar q$ pair in $\gamma_L$ as compared to that in the meson wave function. The same mechanism leads to $Q^2_{eff}/Q^2 \ll 1$  even at large  $Q^2$.

To summarize this section. The 
presence of small size  $q\bar q$ Fock components in light mesons is unambiguously established. At  transverse  separations d $\le$  0.3 fm pQCD reasonably describes Òsmall $"q\bar q$  - dipole" - nucleon interactions for $10^{-4} < x < 10^{-2}$. Color transparency is established for the small dipole interaction  with  nuclei for $x \sim 10^{-2}$. Further  studies of high energy CT and onset of color opacity   will be performed  at LHC  in the  ultraperipheral heavy ion collisions, see \, \cite{Baltz:2007kq}  for a review.

\section{Color transparency for intermediate energies}
\subsection{Expansion effects}
In this section we discuss searches for CT at   Jlab and BNL which correspond to the kinematics where the expansion / contraction of the interacting
small system is very important (essential longitudinal distances are not
 large enough for using of the frozen approximation) and strongly suppresses color transparency effect
\cite{FLFS88,jm90} . 

The maximal longitudinal  distance for  which coherence effects are still 
present is determined by the minimal characteristic internal excitation
energies of the hadron h. The estimates \cite{FLFS88,jm90} show that 
for the case of a nucleon
ejectile
  coherence is completely lost at the distances
$\mbox{l}_c \sim 0.3\div 0.4 \mbox{fm}\cdot  \, \mbox{p}_h$,
where $\mbox{p}_h$ is measured in GeV/c \footnote{It is of interest that a much larger value of $l_c/p_h$ is  assumed in  modeling of heavy ion collisions at RHIC.} 

To describe the effect of the loss of coherence two complementary 
languages were suggested. In Ref.~\cite{FLFS88} based on 
the quark-gluon representation of point-like configuration (PLC) wave function it was argued that the
main effect is quantum diffusion of the wave packet so that  
\begin{eqnarray}
\sigma^{PLC}(Z) =(\sigma_{hard} + {Z\over l_c}[\sigma 
-\sigma_{hard}])
\theta(l_c- Z) +\sigma\theta\left(Z-l_c\right). 
\label{eq:sigdif}\end{eqnarray}
This equation is justified for an early  stage of time development
in the leading logarithmic
approximation when perturbative QCD can
be applied. Also,
one can expect that Eq.~(\ref{eq:sigdif})
smoothly  interpolates between the hard and soft regimes.
A sudden change of $\sigma^{PLC}$ would be inconsistent with
the observation of an early (relatively low Q$^2$)
Bjorken scaling \,  \cite{FS88} .
 Eq.(\ref{eq:sigdif}) implicitly incorporates the
 geometric scaling for the PLC - nucleon interactions which
for the discussed energy range includes nonperturbative
effects.

The time development of the $PLC$ can also be 
obtained by its interaction with a nucleus using  
 a baryonic basis for the wave function of PLC:
 \vspace{-0.5cm}
\begin{eqnarray}
\left| \Psi_{PLC}(t)\right>=\Sigma_{i=1}^{\infty} a_i 
\exp(iE_it)\left| \Psi_{i} \right>=
\nonumber \\
= \exp(iE_1t)\Sigma_{i=1}^{\infty} a_i 
\exp\left({i(m_i^2-m_1^2)t\over 2P}\right)\left| \Psi_{i} \right>,
\end{eqnarray}
where $\left| \Psi_{i} \right>$ are the eigenstates of the Hamiltonian with
masses $m_i$, and $p$ is the momentum of PLC which satisfies 
$E_i \gg m_i$.  As soon as the relative phases of 
the different hadronic components
become large (of the order of one) the coherence is likely to be lost.

Numerical results of the quantum diffusion model \cite{FLFS88,FS88} and the model based on the expansion over hadronic basis with sufficiently large number of intermediate states $^{17c}$
give similar numerical results. However
though
both approaches model 
certain aspects of dynamics of expansion, a 
complete treatment of this phenomenon in QCD is so far missing.
In particular, the 
 phenomenon of spontaneously broken chiral symmetry may
 lead to presence of two scales in the rate of expansion, one 
corresponding to regime where
quarks can be treated as massless, and another 
where virtualities become small enough and quark 
acquire effective masses of the order of 300 MeV.

\subsection{Large angle quasielastic A(p,2p) process}
 
 First data on the CT reaction $A(p,2p)$ were obtained at BNL. They were followed  by the dedicated experiment EVA. The final results of EVA \cite{Aclander:2004zm}
 can be summarized as follows.
 Our calculation within the eikonal approximation  with proper normalization of the wave function agrees well the $\mbox{p}_p$=5.9 GeV/c data. The transparency increases significantly   for $\mbox{p}_p$= 9 GeV/c where $l_c$= 2.7 fm. Hence momenta of the incoming proton $\sim $
10 GeV are  sufficient to rather significantly suppress  expansion effects. Hence one can use proton projectiles with  energies above $\sim $10 GeV to study other aspects of the strong interaction dynamics.
 At the same time eikonal approximation level transparency for $\mbox{p}_p$=11.5 $\div$ 14.2 GeV/c represents a problem for all current models  including those which were specifically suggested to explain initial indications of the non-monotonous
 energy dependence of the transparency. This is because   
 the drop of the transparency occurs over a large range of $s^{\prime}$:    $ \mbox{24 GeV}^2\le s^{\prime}\le  \mbox{30 GeV}^2$ which is  too broad for a resonance \cite{Brodsky:1987xw}  or for interference of quark exchange and Landshoff mechanisms \cite{Ralston:1988rb,Jain:1995dd}\, . 
 
 In any case the trend, if confirmed by future data, would strongly suggest that the leading power  quark exchange mechanism of elastic scattering dominates only at very large energies.  This is consistent with the  recent data from Jlab studies of the large angle Compton scattering. These data  are not described by  the minimal Fock space quark counting rule mechanism,  while they agree well  with predictions based on dominance  of the box diagram contribution   \, \cite{Radyushkin:2004sr,Danagoulian:2007gs} . 
 
 \subsection{Color transparency in meson production}
It is natural to expect that it is easier to reach  CT regime  for the interaction/production  of mesons than for baryons since  only two quarks have to come close together. 

The $J/\psi $ coherent and quasielastic photoproduction experiments did find a weak absorption of $J/\psi$ indicating presence of CT.
There was also  evidence 
for CT  in the $\rho$-meson production. However these experiments did not have good enough resolution in the missing mass to suppress hadron production in the nucleus vertex, making interpretation of these experiments somewhat ambiguous.
 
A high resolution experiment of pion production recently reported 
evidence for the onset of CT \, \cite{:2007gqa} in the process $eA\to e\pi^+ A^*$. In the chosen kinematics $\vec{p}_{\pi}\|\vec {q}$ which minimizes contribution of the elastic rescattering. The coherent length defined as the distance between the point where $\gamma^*$ converted to a $q\bar q$ and where $q\bar q$ interacts with a nucleon - $l_{in}=(Q^2+M^2_{q\bar q}/2q_0)$ is small for the kinematics of \, \cite{:2007gqa} and varies weakly with $Q^2$. This simplifies  interpretation of the $Q^2$ dependence of the transparency as compared to the case of small x where  $l_{in}$ becomes comparable to the nucleus size. The experimental results  agree well with predictions of \cite{Larson:2006ge} where CT was calculated based on the quantum diffusion model - Eq. (\ref{eq:sigdif}). 

It is worth emphasizing also, that  in the Jlab kinematics one probes large x processes, which are dominated  for the pion case (and probably also for the $\rho$-meson case) in the pQCD limit by the contribution of the ERBL region. In this case $l_{in}$ has a different meaning than for small x processes where the DGLAP region dominates. It corresponds to the longitudinal distance between the  point  where $\gamma^*$ knocks out a $q\bar q$ pair from the nucleon and the nucleon center.  This distance can be both positive and negative, and hence its variation does not lead to a change  of the rate of the absorption of the produced pair by the other nucleons.

Results for the $\rho$-meson production where also reported at this workshop \cite{Kawtar}. To interpret this experiment one needs to take into account the effect of absorption due to decays of $\rho^0$ to two pions inside the nucleus, and the elastic rescattering contribution which is more important in this case than in the pion experiment since the data are integrated over a large range of the transverse momenta of the $\rho$ meson \, \cite{FMS07}. Up to these effects,  we  expect similar transparency for this reaction and for $\pi$-meson production. 

\section{Directions for the future studies at Jlab}
There are already approved  plans for extending CT studies of the A(e,e'p),
 A(e,e'$\pi$) reactions to much higher energies at 12 GeV. 
This will finally allow to reach kinematics where $l_c$ is larger than the interaction length for a nucleon/pion in the nuclear media.

A complementary strategy is to use processes where multiple rescatterings dominate in light  nuclei ($^2$H,$^3$He) which allows to suppress the expansion effects. An additional advantage of these processes is that one can use for the calculations generalized eikonal approximation, see review in \cite{Sargsian:2001ax}. In particular,  these reactions are well suited to search for a precursor  of CT - suppression of the configurations in nucleons with  pion cloud in the hard processes like the nucleon form factors at relatively small $Q^2\ge 1 \mbox{GeV}^2$ - chiral transparency 
\cite{Frankfurt:1996ai}. The simplest reaction of this kind is production of a slow $\Delta$ isobar in the process $\mbox{e}^2\mbox{H}\to \mbox{e+p+}\Delta^0$ which should be suppressed in the chiral transparency regime.

Two other examples are  (i) large angle $\gamma + N \to "meson" +N$ reaction in nuclei where one should first look for  
a change of A-dependence from $\propto A^{1/3}$ to $\propto A^{2/3}$ already in the region where expansion effects are large due to transition from the vector dominance regime to the regime of point-like photon interaction in which   photon penetrates to any point in the nucleus, (ii)  A-dependence of virtual compton scattering, namely at what $Q^2$  transition from  vector dominance regime to the CT regime occurs. HERMES data are consistent with our prediction based on CT and closure - but accuracy of the data is moderate.

To summarize, the high energy CT is well established and will be further studied at LHC and EIC.  It is likely that Jlab experiments  at 12 GeV will  observe significant CT effects  for the processes with meson production and will provide allow a  decisive test of whether nucleon form factors at $Q^2\sim 15 \mbox{GeV}^2$ are dominated by PLC or mean field configurations.  CT will allow also to establish interplay between soft and hard physics for many other exclusive large momentum transfer processes at Jlab, EIC, LHC  as well as at  hadronic    factories J-PARC, FAIR.

I thank my collaborators on the studies of CT phenomena for numerous discussions.
\bodymatter\bibliographystyle{ws-procs9x6}
\end{document}